\documentclass[%
 reprint,
superscriptaddress,
 amsmath,amssymb,
 aps,
prx,
nolongbibliography
]{revtex4-2}
\usepackage{placeins}
\usepackage{CJKutf8}
\usepackage{lineno}
\usepackage[colorlinks=true,linkcolor=blue]{hyperref}
\usepackage[squaren]{SIunits}
\usepackage{color}
\usepackage{amsmath}
\usepackage{caption,subcaption}
\usepackage{setspace}
\usepackage{multirow}
\usepackage{threeparttable}
\usepackage{booktabs}
\usepackage{graphicx}
\usepackage{dcolumn}
\usepackage{bm}
\usepackage[T1]{fontenc}

\definecolor{corapink}{rgb}{0.97, 0.51, 0.47}

\definecolor{huablue}{rgb}{0.0, 0.3, 0.6}
\newcommand\si[1]{{\color{huablue} {#1}}}
\newcommand\mpes{\si{S1}}

\newcommand\tslong{\si{S4}}
\newcommand\bla{\si{S5}}

\newcommand\qitaxas{\si{S6}}
\newcommand\fengjianju{\si{S7}}

\begin{document}

\preprint{APS/123-QED}

\title{Mapping Transient Structures of Cyclo[18]Carbon by Computational X-Ray Spectra}

\author{Minrui Wei} 
 \affiliation{MIIT Key Laboratory of Semiconductor Microstructure and Quantum Sensing, School of Physics, Nanjing University of Science and Technology, 210094 Nanjing, China}
 
\author{Sheng-Yu Wang}
 \affiliation{MIIT Key Laboratory of Semiconductor Microstructure and Quantum Sensing, School of Physics, Nanjing University of Science and Technology, 210094 Nanjing, China}
 
\author{Jun-Rong Zhang}
 \affiliation{MIIT Key Laboratory of Semiconductor Microstructure and Quantum Sensing, School of Physics, Nanjing University of Science and Technology, 210094 Nanjing, China}
 
\author{Lu Zhang}
 \affiliation{MIIT Key Laboratory of Semiconductor Microstructure and Quantum Sensing, School of Physics, Nanjing University of Science and Technology, 210094 Nanjing, China}
 
\author{Guoyan Ge}
 \affiliation{MIIT Key Laboratory of Semiconductor Microstructure and Quantum Sensing, School of Physics, Nanjing University of Science and Technology, 210094 Nanjing, China}
 
\author{Zeyu Liu}
\email{liuzy@just.edu.cn}
\affiliation{School of Environmental and Chemical Engineering, Jiangsu University of Science and Technology, 212100 Zhenjiang, China}

\author{Weijie Hua}
 \email{wjhua@njust.edu.cn}
 \affiliation{MIIT Key Laboratory of Semiconductor Microstructure and Quantum Sensing, School of Physics, Nanjing University of Science and Technology, 210094 Nanjing, China}

\date{\today}
\clearpage

\begin{abstract}
The structure of cyclo[18]carbon (C$_{18}$), whether in its polyynic form with bond length alternation (BLA) or its cumulenic form without BLA, has long fascinated researchers, even prior to its successful synthesis. Recent studies suggest a polyynic ground state and a cumulenic transient state; however, the dynamics remain unclear and lack experimental validation. This study presents a first-principles theoretical investigation of the bond lengths ($R_1$ and $R_2$) dependent two-dimensional potential energy surfaces (PESs) of C$_{18}$, concentrating on the ground state and carbon 1s ionized and excited states. We examine the potential of X-ray spectra for determining bond lengths and monitoring transient structures, finding that both X-ray photoelectron (XPS) and absorption (XAS) spectra are sensitive to these variations. Utilizing a library of ground-state minimum structures optimized with 14 different functionals, we observe that core binding energies predicted with the $\omega$B97XD functional can vary by 0.9 eV (290.3--291.2 eV). Unlike the ground state PES, which predicts minima at alternating bond lengths, the C1s ionized state PES predicts minima with equivalent bond lengths. In the XAS spectra, peaks 1$\pi^*$ and 2$\pi^*$ show a redshift with increasing bond lengths along the line where $R_1 = R_2$. Additionally, increasing $R_2$ (with $R_1$ fixed) results in an initial redshift followed by a blueshift, minimizing at $R_1 = R_2$. Major peaks indicate that both 1$\pi^*$ and 2$\pi^*$ arise from two channels: C1s$\rightarrow\pi^*_{z}$ (out-of-plane) and C1s$\rightarrow\pi^*_{xy}$ (in-plane) transitions at coinciding energies.
\end{abstract}

\maketitle
The synthesis of new compounds and the determination of their structures are among the primary objectives of chemical research.\cite{hess_ab_1986} Cyclic carbon molecule is a typical $sp$-hybridized carbon allotrope, which has been the object of interest for a long time. Since Hoffmann\cite{hoffmann_extended_1966} first proposed the structure of C$_{18}$ in 1966, experimental attempts to synthesize it have never stopped.\cite{diederich_all-carbon_1989,rubin_higher_1991, tobe_new_1996}  Many early spectroscopic studies provided evidence for the existence of C$_{18}$ in the gas-phase environment\cite{diederich_all-carbon_1989, diederich_synthetic_1992, diederich_carbon_1994}, but no clear image has been observed.  Until 2019 and 2020, Kaiser $et$ $al$.\cite{kaiser_sp-hybridized_2019} and Scriven $et$ $al$.\cite{scriven_synthesis_2020}  successfully prepared and characterized the C$_{18}$ in the condensed phase, which once again shocked carbon-rich and full-carbon enthusiasts.  

\begin{figure*}[!htb]
  \centering
  \centering\includegraphics[width=17cm]{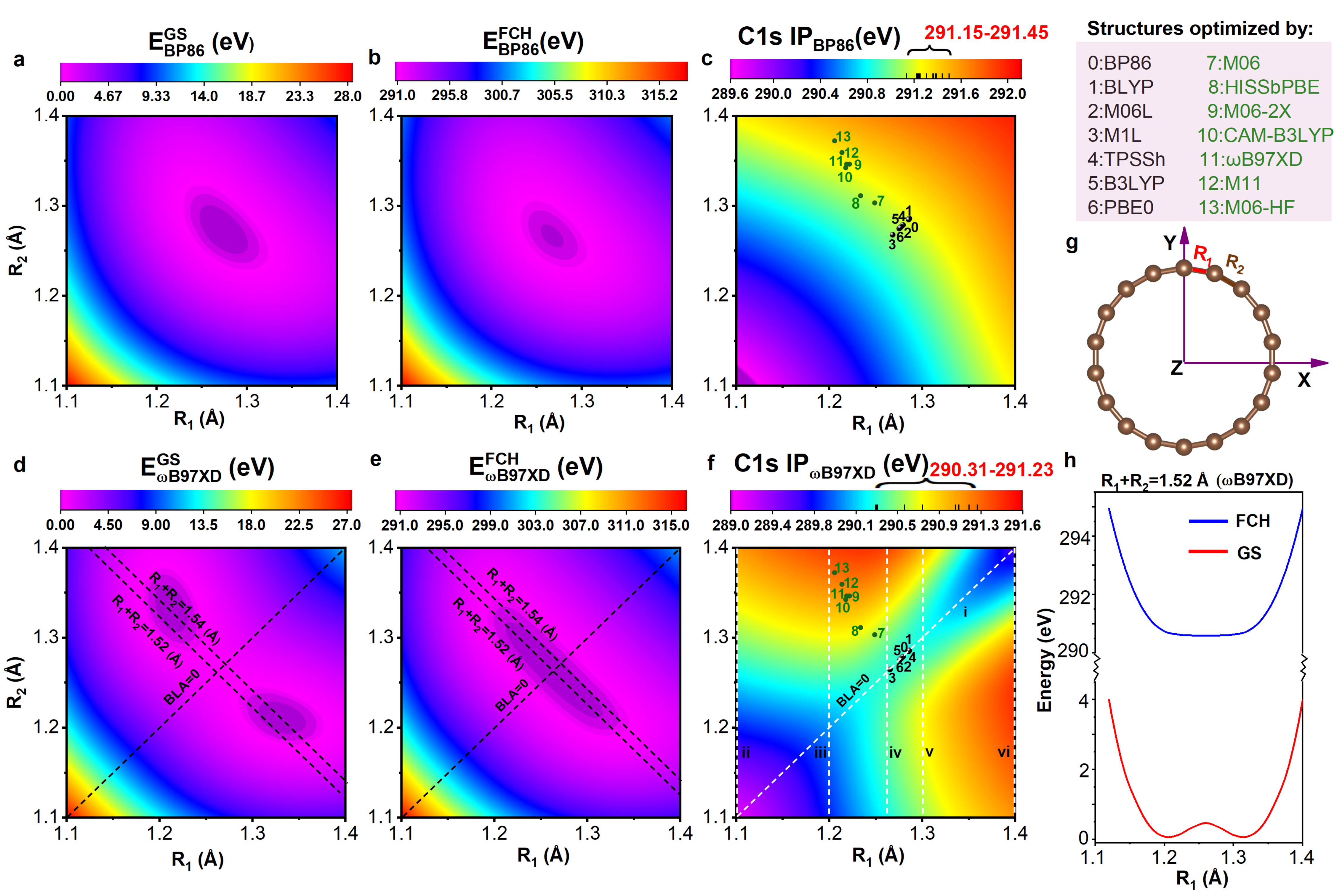}
  \caption{PESs of C$_{18}$  predicted by DFT with two functionals: (a-c) BP86, (d-f) $\omega$B97XD.  (a, d) Ground-state energies  ($E^\text{GS}$), (b, e) C1s-ionized state energies ($E^\text{FCH}$), and (c, f) their difference (i.e., vertical C1s IP).  In panels c and f, indices denote various structures optimized with different functionals. (g) Structure of the C$_{18}$ with $R_1$ and $R_2$ defined. (h) Potential energy curves of the ground and the FCH states along the line $R_1$+$R_2$=1.52 {\AA} calculated using $\omega$B97XD functional.}\label{fig:ip}
\end{figure*}

Before experimental observations, the geometric structure of the C$_{18}$ molecule had been a subject of theoretical debate. Different computational levels predict two types of structures: polyynic and cumulenic. \cite{brzyska_cyclo18carbon_2022}The cumulenic structure with identical C-C bond lengths was determined by theoretical calculations with the MP2.\cite{parasuk_18_1991, feyereisen_relative_1992}  By contrast, Hartree-Fock (HF),\cite{parasuk_18_1991, diederich_all-carbon_1989, ml_martin_structure_1995, plattner_c18_1995} Quantum Monte Carlo (QMC),\cite{torelli_electron_2000} and coupled cluster with singlets and doublets (CCSD)\cite{arulmozhiraja_ccsd_2008, ml_martin_structure_1995} methods supported the polyynic structure with alternating  C-C bonds. For density functional theory (DFT) calculations, the HF exchange in the hybridization function determines the bonding pattern: a cumulenic structure results from functionals with less than 25{\%} HF exchange components (e.g., B3LYP, PBE, PBE0, BLYP),\cite{fowler_double_2009, ott_raman_1998, rechtsteiner_raman_2001, boguslavskiy_gas-phase_2005, murphy_chiroptical_2018, hutter_structures_1994, arulmozhiraja_ccsd_2008, neiss_nature_2014} while a polyyne structure is obtained only when the HF exchange composition of the functional is greater than 25{\%} (e.g., $\omega$B97XD, M06-2X).\cite{stasyuk_cyclo_2020, plattner_c18_1995, liu_sp-hybridized_2020} The latest experiments unambiguously determined that the ground state of C$_{18}$ is a polyyne structure through low-temperature scanning tunneling microscopy and atomic force microscopy (STM-AFM) atomic manipulation techniques, which also identified suitable calculation methods for theoretical research.\cite{kaiser_sp-hybridized_2019, scriven_synthesis_2020} Subsequently, a series of theoretical investigations have been conducted on the properties of C$_{18}$ in condensed\cite{liang_coupling_2020} and gas phases,\cite{lu_accurate_2022,  liu_sp-hybridized_2020, zhang_diverse_2020, dai_achieving_2020, charistos_induced_2020, remya_carbon_2016, liu_vibrational_2020, liu_potential_2022, iyakutti_18_2021, liu_intermolecular_2021, liu_remarkable_2021, wang_bonding_2022, wang_photophysical_2022, suresh_cyclo18carbon_2022} including the structure and energy,\cite{liu_sp-hybridized_bond_2020, lu_accurate_2022} electronic properties,\cite{liu_sp-hybridized_bond_2020, liang_coupling_2020, liu_sp-hybridized_2020, zhang_diverse_2020} aromatics,\cite{liu_sp-hybridized_bond_2020, dai_achieving_2020, charistos_induced_2020} spectral studies (UV-Vis absorption spectrum,\cite{liu_sp-hybridized_2020} IR spectrum,\cite{liu_vibrational_2020} and Raman spectroscopy \cite{liu_vibrational_2020}) intermolecular interactions,\cite{liang_coupling_2020, liu_potential_2022, iyakutti_18_2021, liu_intermolecular_2021} and the properties of related substances\cite{liu_remarkable_2021, wang_bonding_2022, wang_photophysical_2022}, etc.  It was confirmed that C$_{18}$ exhibits extremely high reactivity, giving it great application potential in molecular devices.\cite{zhang_diverse_2020, hou_giant_2021, tang_cross-plane_2023, zhang_c18_2022}

Although the molecular bond type of C$_{18}$ is known, its exact bond lengths have not yet been experimentally determined.\cite{kaiser_sp-hybridized_2019, scriven_synthesis_2020} There are highly advanced imaging techniques in experiments, such as scanning probe microscopy (SPM) based on STM-AFM,\cite{gross_atomic_2018} the high-angle annular dark field (HAADF) scanning transmission electron microscopy
(STEM),\cite{erni_atomic-resolution_2009, sawada_stem_2009} and the electron microscope pixel-array detector (EMPAD). \cite{jiang_electron_2018} However, these methods are mainly used to identify materials or crystal structures, and their resolution is insufficient to determine molecular bond lengths directly.  On the other hand, various spectroscopy techniques play an important role in molecular structure identification.\cite{gross_atomic_2018} The most powerful tools include nuclear magnetic resonance (NMR),\cite{hodgkinson_nmr_2020} mass spectrometry (MS),\cite{samarah_mass_2020, ma_advances_2024} X-ray\cite{tamer_static_2023, ostrom_physisorption-induced_2006, schiros_structure_2006, ge_mapping_2024, ge_qmmm_2022, champenois_femtosecond_2023} and ultraviolet-visible (UV/Vis) spectroscopy,\cite{mcnaughton_machine_2023, tamer_static_2023} which are widely recognized as the gold standard for molecular analysis. Among them, core excitation-based X-ray photoelectron (XPS) and absorption (XAS) spectroscopies are particularly sensitive to the chemical environment, with spectral shapes that can respond sensitively to changes in molecular bond structures. Accurate calculations can readily identify the corresponding structure by comparing them with experimental spectra. Calculated XPS and XAS spectra yield excellent results in determining the excited state bond lengths of diatomic molecules,\cite{neeb_coherent_1994, puttner_vibrationally_1999, chen_k_1989} crystal proton positions,\cite{ge_mapping_2024, ge_qmmm_2022} and molecular structures in the solid state.\cite{ostrom_physisorption-induced_2006, schiros_structure_2006}

This letter aims to explore the applicability of X-ray spectroscopy in determining the bond lengths of C$_{18}$ molecule and monitoring its transient structure. A series of C$_{\text{18}}$ structures in the $x$-$y$ plane were constructed [see Fig. \ref{fig:ip}(e)], including two representative discrete structures with D$_{\text{18h}}$ and D$_{\text{9h}}$ symmetries [Fig. \ref{fig:ip}(g)]. The two adjacent bond lengths, $R_{1}$ and $R_{2}$, were varied from 1.10 to 1.40 {\AA} with an interval of 0.02 {\AA}. By using the DFT method with appropriate $\omega$B97XD functional\cite{chai_long-range_2008}, we comprehensively scanned the bond-length dependent two-dimensional PES of C$_{18}$ at ground state as well as C1s ionization and excited states. Furthermore, for each snapshot structure, XPS and XAS spectra were simulated and analyzed to establish a theoretical reference for future spectroscopic experiments.

The $\Delta$Kohn-Sham scheme\cite{triguero_separate_1999, PhysRev.139.A619} combined with the full core hole (FCH) approximation\cite{zhang_accurate_2019, ge_qmmm_2022} was used to calculate the C1s ionic potential (IP) and calibrate the first transition energy of the XAS spectra. Another effective approximation method, the equivalent core hole method (ECH),\cite{plashkevych_validity_2000, hua_x-ray_2010} whose molecular orbitals (MOs) are used to analyze the transitions in the XAS spectra. Further details of the calculations are given in the supplementary material.\cite{si_C18}

\begin{table}[!htbp]
\centering
\caption{C1s IPs of C$_{18}$ (in eV) simulated at structures optimized with different functionals. At each optimized geometry, bond lengths $R_1$ and $R_2$ [see structure in Fig. \ref{fig:ip}(g)] are given in {\AA}. All IPs ($I$) were computed with the FCH method with the BP86 and $\omega$B97XD functionals, respectively.}
\scalebox{0.8}{
\label{tab:functionals}
\begin{tabular}{lllcccc}
\toprule
Index & Structure &Symmetry&$R_1$& $R_2$& $I_\text{BP86}$& $I_{\omega\text{B97XD}}$\cr
\midrule
        0 &\textbf{min M11L}&D$_\text{18h}$&1.265&1.265&291.04 &290.33\cr
        1 &\textbf{min M06L}&D$_\text{18h}$&1.275&1.275&291.12 &290.33\cr
        2 &\textbf{min PBE0}&D$_\text{18h}$&1.276&1.276&291.13 &290.32\cr
        3 &\textbf{min B3LYP}&D$_\text{18h}$&1.277&1.277&291.13 &290.32\cr
        4 &\textbf{min TPSSh}&D$_\text{18h}$&1.279&1.279&291.15 &290.32\cr
        5 &\textbf{min BP86}&D$_\text{18h}$&1.285&1.285& 291.20 &290.31\cr
        6 &\textbf{min BLYP}&D$_\text{18h}$&1.286&1.286& 291.21 &290.31\cr
        7 &\textbf{min HISSbPBE}&D$_\text{9h}$&1.235&1.310&291.14 &290.73\cr
        8 &\textbf{min M06}&D$_\text{9h}$&1.250&1.302&291.15 &290.57\cr
        9 &\textbf{min CAM-B3YLP}&D$_\text{9h}$&1.219&1.341&291.26 &291.03\cr
        10 &\textbf{min $\boldsymbol{\omega}$B97XD} &D$_\text{9h}$&1.220 &1.345&291.28 &291.06\cr
        11 &\textbf{min M06-2X}&D$_\text{9h}$&1.223&1.345&291.29 &291.06\cr
        12 &\textbf{min M11}&D$_\text{9h}$& 1.215 &1.358&291.34 &291.15\cr
        13 &\textbf{min M06-HF}&D$_\text{9h}$&1.207&1.371 &291.40 &291.23\cr
\bottomrule
\end{tabular}
\label{tab:14func}
}
\end{table}

Figure \ref{fig:ip} displays the simulated PESs of the ground and core-ionized states of C$_{18}$ by DFT against distances $R_1$ and $R_2$, together with their difference (i.e., vertical IPs). Two different functionals,  the pure generalized gradient approximation (GGA) functional (BP86) and the range separated hybrid GGA functional ($\omega$B97XD) were used, revealing two distinct scenarios for the ground-state energies. A single minimum [Fig. \ref{fig:ip}(a)] and dual minima [Fig. \ref{fig:ip}(d)] were identified, respectively.  The BP86 PES [Fig. \ref{fig:ip}(a)] covers an expansive energy spectrum of 28.4 eV, wherein the cumulenic structure with $R_{1}$=$R_{2}$=1.28 {\AA} is pinpointed as the point of minimal energy.  By contrast, the $\omega$B97XD functional shows a maximum energy difference of 27.4 eV and identifies minima at polyynic structures with ($R_{1}$,  $R_{2}$) values at (1.22, 1.32) and (1.32, 1.22) {\AA}, respectively (the combinational notation was used hereafter).  Previous GS calculations showed that the $\omega$B97XD  functional predicts more accurate structures.\cite{liu_sp-hybridized_bond_2020, liu_comment_2021} Additionally, calculations using the M06-2X functional yield similar results to $\omega$B97XD [Fig. {\mpes}(a)].

Regarding the PESs in the core-ionized state, there is a notable discrepancy between results obtained by using the BP86 and $\omega$B97XD functionals.  Figure \ref{fig:ip}(b) illustrates that the BP86 PES exhibits only a single minimum at (1.28, 1.28) {\AA}. Conversely, in the $\omega$B97XD PES, a distinct oval-shaped low energy band is observed in the center region, with the longer axis staying at $R_1+R_2=1.52$ {\AA} and the shorter axis at $R_1=R_2$. Along the long axis [Fig. \ref{fig:ip}(h)], the potential energy curve shows a flat bottom region. Upon core ionization, the evident double-well feature in the ground-state PEC becomes much weaker, and it becomes difficult to distinguish the two minima within the flat bottom. Sliced PECs along the long axis and an additional parallel line $R_1+R_2=1.54$ are depicted in Fig. {\tslong} (a)-(b), clearly illustrating the flat band features. Meanwhile, along the short axis, PECs of both the ground and core-ionized states exhibit similar parabolic curves, with their minima both identified at (1.26, 1.26) {\AA} [Fig. {\bla}(a)].  Additional calculations were performed with the M06-2X functional, and consistent results with $\omega$B97XD were achieved (Figs. {\tslong}-{\bla}). Our results indicate significant changes in the PES as induced by core ionization. 

Furthermore, Fig. \ref{fig:ip}(c) and \ref{fig:ip}(f) depict the contour maps for the vertical ionic potential (i.e., IP value, the energy difference between the core-ionized and the ground states) simulated by  BP86 and $\omega$B97XD functionals, respectively. The two 2D maps exhibit distinctly different features, with maximum chemical shifts of 2.48 and 2.35 eV, respectively. The IP values calculated by the BP86 function exhibit a direct monotonic increase as the bond lengths increase [Fig. \ref{fig:ip}(c)]. While the $\omega$B97XD functional produces a more complex pattern [Fig. \ref{fig:ip}(f)], demonstrating the sensitivity of the C1s  ionic potential to variations in the bond lengths of the C$_{18}$ molecule.  In general, two high-energy regions (in red color) appear centered at around (1.24, 1.40) and (1.40, 1.24) {\AA}. Interestingly, low energy regions appear in both small and large distances (in blue and purple colors) along the $R_1=R_2$ line, resulting from the cancellation of potential energies between the two states. For a series of 14 ground-state structures\cite{stasyuk_cyclo_2020} relaxed using different functionals (Table \ref{tab:14func}, the $\omega$B97XD functional predicts a C1s IP range of 0.9 eV (290.31-291.23 eV), which is significantly higher than the 0.3 eV (290.31-291.23 eV) simulated by the BP86 functional. In addition, the $\omega$B97XD results distinctly demarcate the D$_\text{18h}$ and D$_\text{9h}$ structures, segregating them into respective low- (290.31--290.33 eV) and high-energy (290.73--291.23 eV) regions. Therefore, based on the correct simulation method, C1s IP values are highly sensitive to response to variations in bond lengths. 

\begin{figure*}[!htbp]
  \centering
  \includegraphics[width=16.2cm]{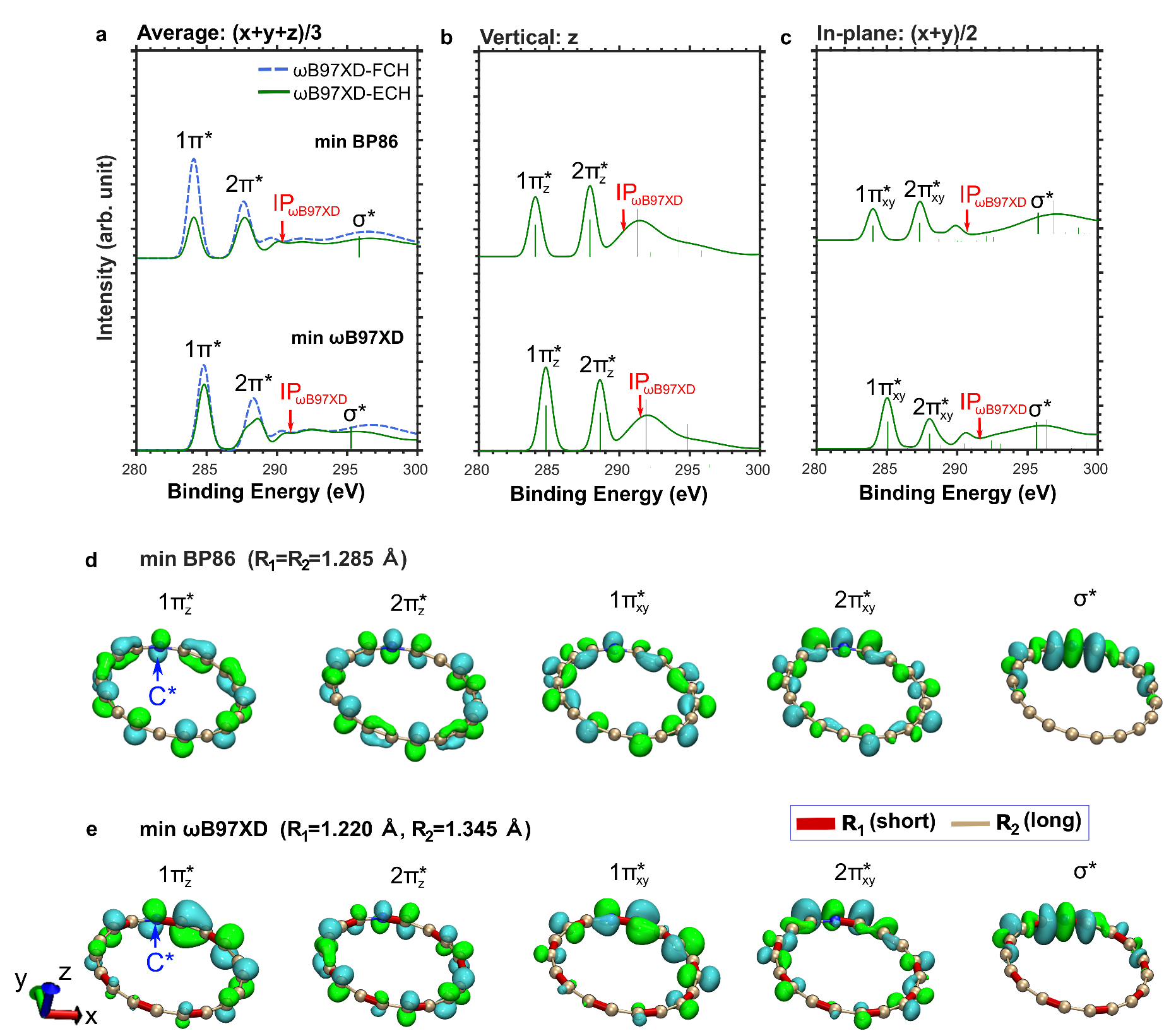}
  \caption{Analysis of simulated C1s NEXAFS spectra of C$_{18}$. Total spectra calculated by the FCH and ECH methods with the $\omega$B97XD functional at cumulenic and polyynic structures. (a) Comparison of total spectra by ECH and FCH methods.  (b-c) 1s$\rightarrow\pi^*$ transitions. (d-e) Final-state MOs for major transitions in panels b and c at the two structures.  Contour isovalue{=}0.02 was used. The blue arrow denotes the position of the excited carbon. } 
  \label{fig:orbital}
\end{figure*}

Based on the C$_{18}$ structures with different bond lengths, we simulated the XAS spectra at various snapshots with $\omega$B97XD functional. First, XAS spectral analysis was employed to differentiate the molecular configurations exhibiting cumulenic and polyynic structures.  Research indicates that the cumulenic geometry might be a transition state structure for the bond transfer between the two polyyne forms (reversed order of single and triple bonds) of the C$_{18}$.\cite{baryshnikov_cyclo_2019} Figure \ref{fig:orbital} displays the simulated XAS spectra of C$_{18}$ with cumulene-type geometry (\textbf{min BP86} with D$_{18h}$ symmetry, $R_1$=$R_2$=1.285 {\AA}) and polyyne forms [\textbf{min $\omega$B97XD} with D$_{9h}$ symmetry, ($R_1$, $R_2$)=(1.220, 1.345) {\AA}]  by using two different methods, FCH and ECH.  The FCH approximation is widely recognized for its ability to provide accurate X-ray spectra, but its MOs are unsuitable for analyzing transitions. To better understand the characteristics of XAS, we adopted the ECH method, which performs well in analyzing K-edge XAS of molecules,\cite{hua_x-ray_2010} crystals,\cite{ge_qmmm_2022}, and 2D materials.\cite{zhang_accurate_2019}

Figure \ref{fig:orbital}(a) depicts the precise spectra of cumulenic and polyynic structures calculated by the $\omega$B97XD-FCH method. The polyynic structure exhibits a noticeable blue shift compared to the cumulenic structure. Through ECH analysis, two $\pi^*$ peaks are exhibited in both spectra, with a strong 1$\pi^*$ peak and a weaker 2$\pi^*$ peak, along with a discernible $\sigma^*$ feature at the tail end of the spectra.  Among them,  1(2)$\pi^*$ peak is contributed by the 1(2)$\pi_\text{xy}^*$ [Fig. \ref{fig:orbital}(b)] and 1(2)$\pi_\text{z}^*$ [Fig. \ref{fig:orbital}(c)]. Figure \ref{fig:orbital}(d-e) presents the MOs analysis of each characteristic peak as depicted in Fig. \ref{fig:orbital}(b-c). MOs at the $\pi$ peaks in the two configurations exhibit significant differences. For cumulenic structure [Fig. \ref{fig:orbital}(d)] , MOs at the four $\pi$ peaks demonstrate good symmetry about the y-axis.  MO at the low-energy peak 1$\pi_\text{xy}^*$ (1$\pi_\text{z}^*$) is predominantly distributed on the bonds surrounding the core hole, whereas for the high-energy peak 2$\pi_\text{xy}^*$ (2$\pi_\text{z}^*$), the MO is primarily localized on the atoms.   Compared to the cumulenic structure, the MOs of the $\pi$ polyynic structure [Fig. \ref{fig:orbital}(e)] no longer exhibit axial symmetry and behave more localized. MO at the low energy peak 1$\pi_\text{xy}^*$ (1$\pi_\text{z}^*$) are stacked on short bonds near the core hole, while the MO at the high energy peak 2$\pi_\text{xy}^*$ (2$\pi_\text{z}^*$) spreads to the atoms.  Furthermore, we analyzed the MOs at the $\sigma^*$ characteristics of both configurations and observed similar behavior, consistently localized around the core hole.

\begin{figure}[!htb]
  \centering
  \includegraphics[width=8.6cm]{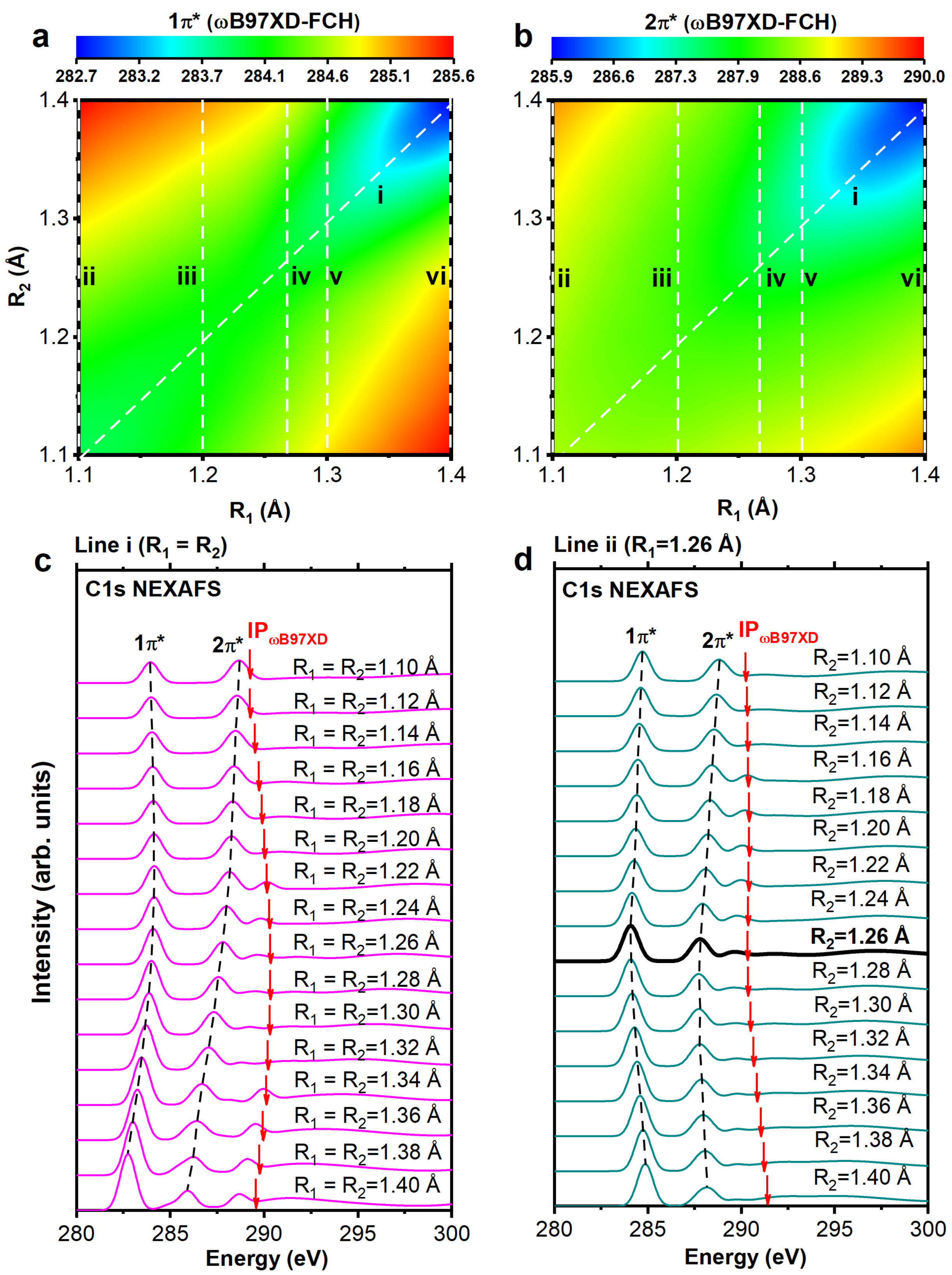}
  \caption{Computed C1s NEXAFS spectra of of C$_{18}$. 
(a-b) Energy contour plot of peaks (a) 1${\pi}^*$ and (b) 2${\pi}^*$ computed by the $\omega$B97XD-FCH method. (c-d) Total NEXAFS spectra along two lines in panels a--b: (c) the diagonal line i with $R_1=R_2$; (d) the vertical line iii with fixed bond length $R_1$=1.26 {\AA}. Spectra of lines ii, iv, v and vi are provided in Fig. {\qitaxas}(a--d). }
  \label{fig:xas}
\end{figure}

Figure \ref{fig:xas}(a)-(b) displays the 2D maps of the 1${\pi}^*$ and 2${\pi}^*$ energy positions at varying bond length configurations. Varying bond lengths cause a chemical shift of ca. 4 eV for both the 1${\pi}^*$ and 2${\pi}^*$ energies, spanning ranges of 282–286 eV and 286–290 eV, respectively, indicating a significant energy-structure dependence. To thoroughly examine the effects of alterations in bond length, we sliced the two PESs along six lines, encompassing a diagonal line (line i, $R_1$=$R_2$) and five lines with a fixed $R_1$ bond length (lines ii-vi, $R_1$ is fixed to 1.10, 1.20, 1.26, 1.30, 1.40 {\AA}, respectively). Along $R_1$=$R_2$ line [Fig. \ref{fig:xas}(c)], increasing bond length from 1.10 to 1.40 {\AA} results in a blue shift of 1${\pi}^*$ (2${\pi}^*$) energy by ca. 1.2 (2.7) eV.  However, for the lines with keeping $R_1$ fixed [Figs. \ref{fig:xas}(d) and {\qitaxas}(a-d)], the responses of 1${\pi}^*$ and 2${\pi}^*$ energies to the bond length change are more complicated. The energy minima of 1${\pi}^*$ and 2${\pi}^*$ are consistently located at $R_1$=$R_2$.  The energy shifts of the two characteristic peaks can be divided into two stages: when $R_1$ $\textgreater$ $R_2$, increasing $R_2$ causes the energy of 1${\pi}^*$ and 2${\pi}^*$ to gradually redshift, while when $R_1$ $\textless$ $R_2$, it causes a blueshift. 

Additionally, the separation $\delta$ between the 1${\pi}^*$ and 2${\pi}^*$ peaks of the structures along the six defined lines is analyzed (Fig. {\fengjianju}), as it is an important parameter for comparison with experimental spectra.\cite{ge_mapping_2024} When $R_1 = R_2$ [Fig. {\fengjianju}(a)], increasing the bond length from 1.10 to 1.40 {\AA} leads to a gradual decrease of $\delta$ by 1.9 eV (from 1.4 to 3.3 eV). For a constant $R_1$ bond length, increasing $R_2$ within the range of 1.1 to 1.4 {\AA} monotonously reduces $\delta$. Specifically, as $R_1$ takes on values of 1.1, 1.2, 1.26, 1.3, and 1.4 {\AA}, the observed decreases in peak separation are 0.93, 0.83, 0.85, 0.68, and 0.65 eV, respectively [Fig. {\fengjianju}(b-f)].

In summary, we have utilized DFT to generate bond length-dependent 2D potential energy surfaces (PESs) of C$_{18}$ in both ground and C1s ionized/excited states, exploring the applicability of X-ray spectra for determining bond lengths and transient intermediate structures. Our findings reveal that XPS and XAS spectra are highly sensitive to variations in bond lengths, serving as effective probes for dynamic changes. Using the $\omega$B97XD or M06-2X functional, the ground-state PES exhibits two minima with alternating bond lengths. Notably, in the core-excited state PES, the minima correspond to a cumulenic structure with equivalent bond lengths, resulting in chemical shifts spanning 2.38 eV across the map. Analysis of ground-state minimum structures optimized with 14 different functionals shows that core binding energies predicted by the $\omega$B97XD functional can vary by 0.9 eV (from 290.3 to 291.2 eV).

The XAS simulation identifies two $\pi^*$ peaks that are sensitive to bond length changes. For configurations where $R_1 = R_2$, increasing the bond length leads to a gradual redshift of the 1$\pi^*$ and 2$\pi^*$ peaks, with the separation $\delta$ changing from 1.4 to 3.3 eV, reflecting a shift of 1.9 eV. When $R_1$ is fixed, increasing $R_2$ results in both peaks exhibiting a redshift followed by a blueshift, reaching a minimum when $R_1 = R_2$. Meanwhile, $\delta$ decreases monotonically by 0.7–0.9 eV. The main peaks are attributed to $\pi^*$ contributions from both $\pi_\text{z}^*$ and $\pi_\text{xy}^*$ orbitals, with an additional $\sigma^*$ feature identified in the tail's transition region. Final-state molecular orbital analysis indicates that the $\sigma^*$ orbital is more delocalized than the $\pi^*$ orbitals. This computational spectroscopy study establishes a structure-spectroscopy relationship that can inform future experimental investigations into dynamic processes.

\begin{acknowledgments}
Financial support from the National Natural Science Foundation of China (Grant No. 12274229) is greatly acknowledged.
\end{acknowledgments} 

\FloatBarrier

%

\end{document}